\documentclass[twocolumn,aps,groupedaddress]{revtex4}

\usepackage{graphicx}
\usepackage{dcolumn}
\usepackage{bm}
\usepackage{amsmath,amssymb,amsfonts}
\usepackage{float}
\usepackage{color}
\usepackage{dcolumn}
\usepackage{wasysym}
\newcolumntype{N}{@{}m{0pt}@{}}

\begin{document}

\title{Moir\'e quantum chemistry: charge transfer in transition metal dichalcogenide superlattices}

\author{Yang Zhang}\thanks{These two authors contributed equally to this work.}
\author{Noah F. Q. Yuan}\thanks{These two authors contributed equally to this work.}
\author{Liang Fu}
\affiliation{Department of Physics, Massachusetts Institute of Technology, Cambridge, Massachusetts 02139, USA}

\begin{abstract}
Transition metal dichalcogenide (TMD) bilayers have recently emerged as a robust and tunable moir\'e system for studying and designing correlated electron physics. In this work, by combining large-scale first principle calculation and continuum model approach, we provide an electronic structure theory that maps long-period heterobilayer TMD superlattices onto diatomic crystals with cations and anions. We find that the interplay between moir\'e potential and Coulomb interaction leads to filling-dependent charge transfer between
MM and MX regions several nanometers apart. We show that the insulating state at half-filling found in recent experiments on WSe$_2$/WS$_2$ is a charge-transfer insulator rather than a Mott-Hubbard insulator. Our work reveals the richness of simplicity in moir\'e quantum chemistry.
\end{abstract}

\maketitle

Following the recent discovery of correlated insulators and unconventional superconductivity in twisted bilayer graphene \cite{cao2018correlated,cao2018unconventional} and trilayer graphene-hBN heterostructure \cite{chen2019evidence,chen2019sig}, artificial moir\'e superlattices have emerged as a new venue for realizing and controlling correlated electron phenomena.
The moir\'e superlattices and natural solids differ greatly in the magnitude of characteristic length and energy. In solids, the average distance between electrons is typically comparable to atomic spacing in the order of ${\rm\AA}$ and their kinetic and interaction energies are typically in the order of eV, while in moir\'e superlattices a mobile charge is shared by 1000--10000 atoms so that the characteristic length and energy scales are in the order of 10 nm and 10--100 meV respectively. Correspondingly, the quantum chemistry of natural solids involves complex intra-atomic and long-range interactions, while low-energy charge carriers in moir\'e superlattices only feel a long-period potential and interact with each other predominantly via the long-range Coulomb repulsion. Therefore, quantum chemistry can be simpler in moir\'e systems.

In twisted bilayer graphene, the emergence of strong correlation effects requires fine tuning to a magic twist angle, where the moir\'e energy bands become flattened \cite{bistritzer2011moire} and sensitive to microscopic details such as lattice relaxation \cite{nam2017lattice,koshino2018maximally,carr2019exact,yoo2019atomic,uchida2014atomic,van2015relaxation} and strain \cite{bi2019designing}.
On the other hand, transition metal dichalcogenide (TMD) bilayers \cite{wu2018hubbard,wu2019topological,tang2019wse2,regan2019optical} have a much simpler moir\'e band structure. In TMD heterobilayers such as WSe$_2$/WS$_2$, the valence moir\'e bands are simply formed by holes moving in a periodic moir\'e potential. Therefore, TMD superlattices provide a robust platform to study many-body physics with a highly tunable kinetic energy and local interaction strength.

Very recently, a correlated insulating phase has been observed in WSe$_2$/WS$_2$ at half-filling \cite{tang2019wse2,regan2019optical} of the topmost valence moir\'e bands with a charge gap around 150K ($\sim$ 10 meV) , and regarded as a canonical Mott-Hubbard insulator \cite{hubbard1963electron}. In this scenario, the topmost moir\'e band is well separated from the rest; its charge distribution is tightly localized near the moir\'e potential minima, forming a triangular lattice. Strong on-site Coulomb repulsion $U$ suppresses double occupancy and creates an insulating gap at half-filling in the order of $U$. 

In this work, we identify a new energy scale associated with charge transfer between regions with different local stacking configurations in the moir\'e superlattice. When the energy cost of charge transfer $\Delta$ is comparable to or smaller than the local Coulomb repulsion $U$, the Mott-Hubbard description becomes inadequate. Instead, we show that a new type of correlated insulator emerges at half-filling, known as the charge-transfer insulator \cite{zaanen1985band}. Using large-scale first-principles calculation, we obtain the parameters $\Delta$ for various TMD heterobilayers and find that $\Delta$ in WSe$_2$/WS$_2$ superlattice is comparable to the experimentally observed charge gap\cite{tang2019wse2,regan2019optical}, whereas $U$ is much larger. We provide a theoretical description of the charge transfer phenomenon by introducing an effective honeycomb lattice model, in which the $MM$ and $MX$ sublattices correspond to ``moir\'e cations'' and ``moir\'e anions'' where charges are locally concentrated. 
We note that previous works on twisted bilayer graphene have shown the interaction induced charge redistribution within a local moire region\cite{rademaker2018charge,rademaker2019charge,guinea2018electrostatic}. Here the charge transfer we predict in TMD moir\'e superlattices takes place on the length scale of the moir\'e period ($\sim$ 10 nm) and can be directly observed by scanning tunneling spectroscopy (STS).

\begin{figure}[t]
\includegraphics[width=1.0\linewidth]{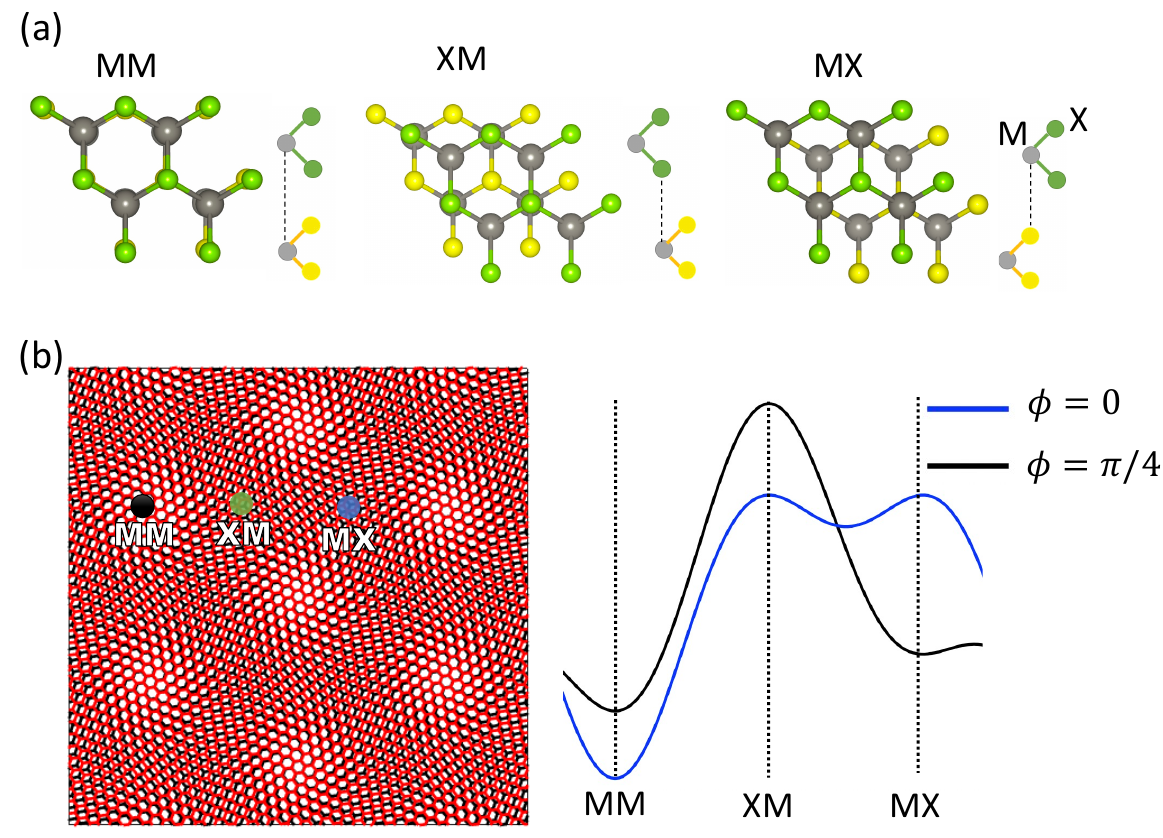}
\caption{
(a) Lattice structure of MM, MX, XM spots for AA stacking heterobilayer (M stands for metal atom and X stands for chalcogen atom),
(b) Real-space moir\'e pattern of heterobilayer TMD heterobilayer with $\delta=4.0\%$, where MM, MX, XM spots within one supercell are labeled, and schematic diagram for moir\'e potential landscape for $\phi=0$ and $\phi=\pi/4$, along the path from MM to XM and MX spots as indicated by the array in the left figure.
}
\label{fig1}
\end{figure}

\begin{figure}[t]
\includegraphics[width=1.0\linewidth]{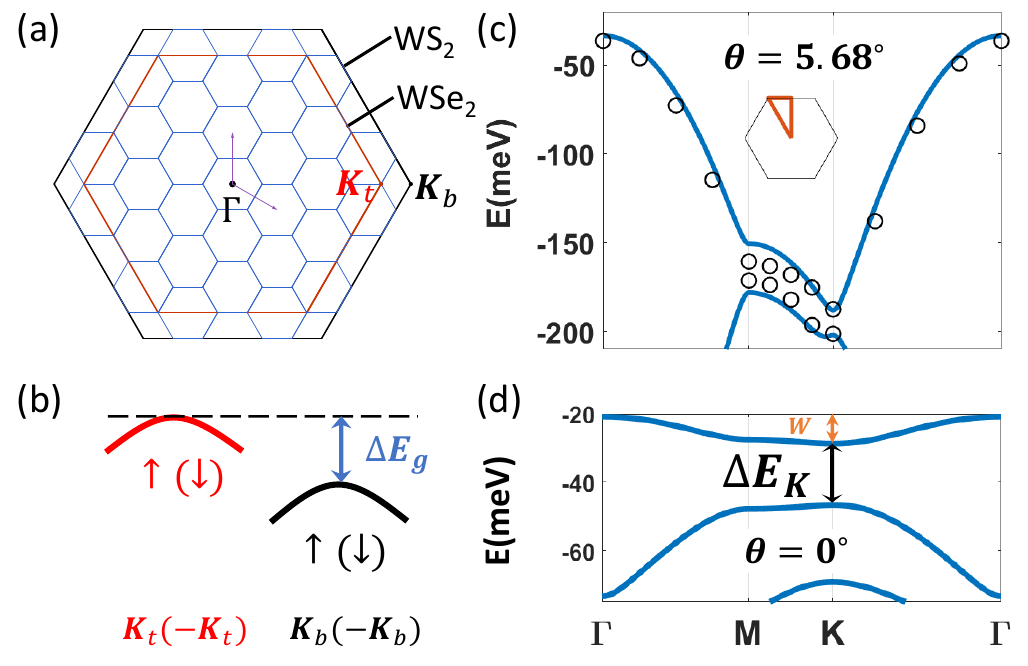}
\caption{
(a) Brillouin zone (BZ) folding in WSe$_2$/WS$_2$
moir\'e superlattice, where $a_{WSe_2}/a_{WS_2}$ is taken as 6/5 for the illustrative purpose.
(b) Schematic low-energy band structure from two layers where $\pm\bm K_{t(b)}$ are two valleys of top (bottom) layer. 
(c) DFT band structure(open circle) and continuum model band structure (blue lines) at $\theta=5.68^\circ$,
(d) continuum model band structures of WSe$_2$/WS$_2$ at $\theta=0^\circ$.
}
\label{fig2}
\end{figure}

\emph{Continuum Model}---
We consider a heterobilayer TMD such as WSe$_2$/WS$_2$, with $a(a')$ as the lattice constant of top (bottom) layer, and $\theta$ as the twist angle. The lattice mismatch leads to a moir\'e superlattice in Fig. \ref{fig1}, with superlattice constant $L_{\rm M}=a/\sqrt{\delta^2+\theta^2}$ where $\delta=(a-a')/a'$. As illustrated in Fig. \ref{fig2}b, the valence bands of two layers have a large band offset $\Delta E_g$, which is listed for various TMD heterobilayers at zero twist angle in Table. \ref{T}. 
Given the large band offset, the low-energy moir\'e bands result from the spatial variation of valence band maximum of WSe$_2$ due to the lattice mismatch with WS$_2$ \cite{zhang2017interlayer}, which is described by a long-period moir\'e potential acting on holes in WSe$_2$.

In this work, we study TMD heterobilayers with a small twist angle starting from AA stacking, where the metal atom and chalcogen atom of the top layer are aligned with metal atom and chalcogen atom at the bottom layer, respectively\footnote{Alternatively, AB stacking can be viewed as a $180^{\circ}$ rotation of top layer}. There are three types of Wykoff positions in a moir\'e unit cell---hereafter referred to as MM, XM, MX, depending on the alignment of the metal atom (M) and chalcogen atom (X). As shown in Fig. \ref{fig1}a, at MM, the metal atoms on top and bottom layers are aligned, 
While at MX (XM), the metal atom on the top (bottom) layer is aligned with the chalcogen atom on the bottom (top) layer. In the long moir\'e wavelength limit $L_{\rm M}/a\to\infty$, the valence band maximum varies slowly over the moir\'e unit cell, which can be expressed as the first-order harmonics with  moir\'e wave vectors $\bm G_i=\frac{4\pi}{\sqrt{3}}L_{\rm M}^{-1}(\cos\frac{i2\pi}{3},\sin\frac{i2\pi}{3})$ (i=1,2,3)\cite{wu2018hubbard}.

This is captured by the continuum model  $H_0=\int\psi^{\dagger}(\bm r)\hat{\mathcal{H}}\psi(\bm r)d^2\bm r$ with
\begin{eqnarray}\label{eq_cm}
\hat{\mathcal{H}}&=&-\frac{\nabla^2}{2m}+V(\bm r),\\\label{eq_v}
V(\bm r)&=&-2V_{0}\sum_{i=1}^{3}\cos(\bm G_{i}\cdot\bm r+\phi),
\end{eqnarray}
where $\psi^\dagger=(\psi^\dagger_{\uparrow},\psi^\dagger_{\downarrow})$ creates the holes and $m>0$ is the effective mass. From first principle calculation with the relaxed layer spacing(see Supplemental Material), we find the
moir\'e valence bands within 200 meV are formed by $\pm K$ pockets in WSe$_2$.
Owing to strong Ising spin-orbit coupling, valley indices are locked with spin \citep{xiao2012coupled}, and the two valleys are decoupled at long moir\'e wavelength limit.  $V_0>0$ and $\phi$ are the only parameters associated with the magnitude and overall phase of the three lowest Fourier components of the moir\'e potential. When the moir\'e period is large, $V_0,\phi$ are intrinsic material properties independent of $L_{\rm M}$, which we hereafter refer to as moir\'e potential strength and moir\'e phase respectively.

\begin{table}\setlength{\tabcolsep}{7pt}
\centering
\begin{tabular}{c c c c c c}
\hline\hline
System & $\delta$ & $\displaystyle \Delta E_g$ & $\displaystyle V_0$ & $\displaystyle \phi$ & $\displaystyle E_{0}^{\rm min}$ \\[0.3cm]
\hline
WSe$_2$/WS$_2$ & 4\% & 640 & 15 & 45$^\circ$ & 1.2\\[0.2cm]
WSe$_2$/MoS$_2$ & 4\% & 940 & 11 & 40$^\circ$ & 1.2\\[0.2cm]
MoSe$_2$/MoS$_2$ & 4\% & 630 & 9 & 42$^\circ$ & 1.3\\[0.2cm]
MoSe$_2$/WS$_2$ & 4\% & 270 & 7 & 35$^\circ$ & 1.3\\[0.2cm]
\hline\hline
\end{tabular}
\caption{Summary of heterobilayer TMD. Here $\delta$ is the lattice constant mismatch with respect to the bottom layer, $\Delta E_g$ is the band offset, $V_0$ and $\phi$ are parameters of moir\'e potential and $E_{0}^{\rm min}=\delta^2/(2ma^2)$ is the moir\'e kinetic energy at zero twist. All energies are in unit of meV.}
\label{T}
\end{table}

\begin{figure}[t]
\includegraphics[width=0.85\linewidth]{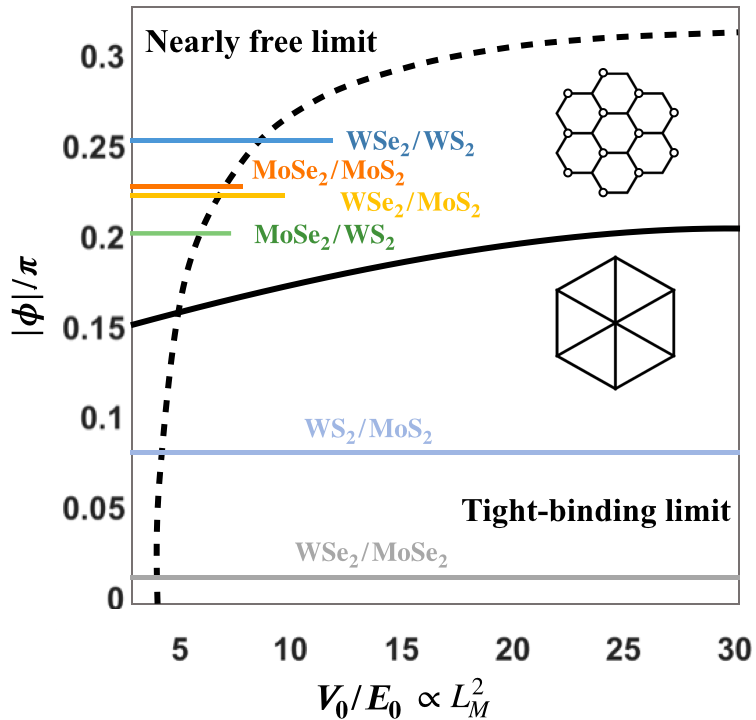}
\caption{Systems listed in Table. \ref{T} and WSe$_2$/MoSe$_2$, WS$_2$/MoS$_2$ (model parameters in Supplementary Material) can be described by different limits and tight-binding models for the first and second moir\'e bands. Each colored line denotes a bilayer TMD, and ends at untwisted limit where $V_0/E_0$ reaches maximum $V_0/E_0^{\rm min}$.
Nearly free limit and tight-binding limit are separated by the dashed line where $W=\Delta$. Within the tight-binding limit, the tight-binding model is a honeycomb lattice formed by MM and MX (open circle) spots with $s$ orbital on each site above the solid black line, and a triangular lattice formed by MM spots with $s,(p_x,p_y)$ orbitals on each site below the solid black line.
}
\label{fig3}
\end{figure}

To obtain the values of $V_0, \phi$, we first use the large-scale density functional theory (DFT) to calculate the moir\'e band structure of WSe$_2$/WS$_2$, WSe$_2$/MoS$_2$, MoSe$_2$/MoS$_2$, and MoSe$_2$/WS$_2$ at the commensurate angel $\theta=5.68^{\circ}$, as shown in Fig. \ref{fig2}c. The spin-orbit coupling is included via DFT\cite{naik2018ultraflatbands,xian2020realization} as implemented in the Vienna Ab \textit{initio} Simulation Package\cite{kresse1996efficiency}. The interaction between electrons and ionic cores is approximated by the projector augmented wave method, and the exchange-correlation potential was described by the Perdew-Burke-Ernzerhof generalized gradient approximation\cite{perdew1996generalized} with the vdW correction incorporated by the vdW-DF (optB86) functionals \cite{klimevs2011van}. We assume rigid lattice along with in-plane directions and relax interlayer distance. Depending on the different vdW correction methods, the interlayer spacing is $6.57\sim6.77$ Angstrom. Throughout this range of interlayer distance, we find the moir\'e band structure is nearly identical.

We find the DFT band structure fits nicely with the continuum model (see Fig. \ref{fig2}c), and obtain from this fitting the material specific parameters $V_0$ and $\phi$ shown in Table. \ref{T}. For WSe$_2$/WS$_2$,
$V_0 = 15$ meV and $\phi = \frac{1}{4}\pi$. 
Importantly, the moir\'e phase $\phi$ determines the energy landscape of moir\'e potential. This can be seen from $V(\bm r)$ at three $C_3$-symmetric points (Wyckoff positions) $\bm r_{\rm MM}=\bm 0,\quad\bm r_{\rm MX}=\frac{1}{\sqrt{3}}L_{\rm M}(1,0)$ and $\bm r_{\rm XM}=-\bm r_{\rm MX}$ respectively.
For $0<\phi<\frac{1}{6}\pi$, within one supercell there are one potential minimum (MM) and two maxima (MX and XM), while for $\frac{1}{6}\pi<\phi<\frac{1}{3}\pi$, there are two minima (MM and MX) and one maximum (XM). The four TMD heterobilayers listed in Table. \ref{T}, WSe$_2$/WS$_2$, WSe$_2$/MoS$_2$, MoSe$_2$/MoS$_2$, and MoSe$_2$/WS$_2$, all belong to the parameter range $\frac{1}{6}\pi<\phi<\frac{1}{3}\pi$, where the presence of two potential minima introduces new physics as we shall show below\footnote{WS$_2$/MoS$_2$ and WSe$_2$/MoSe$_2$ are discussed in Supplementary material.}.

In the following sections, we will study interaction effects in TMD heterobilayers in various regimes of $V_0$ and $\phi$. We denote $n_{s}=2$ holes per supercell as the full filling and $n=\frac{1}{2}n_s=1$ hole per supercell as the half-filling.

\emph{Charge-Transfer Phenomena}---
In this section, we use Hartree approximation to study the effect of Coulomb interaction on the charge distribution in twisted heterobilayer TMD with relatively large bandwidth and demonstrate the charge transfer phenomenon.

The Coulomb interaction including background effect is
\begin{equation}
 H_{C}=\int\delta\rho(\bm r)C(\bm r-\bm r')\delta\rho(\bm r')d^2\bm rd^2\bm r'
\end{equation}
where $\delta\rho\equiv\psi^{\dagger}\psi-\overline{\rho}$ is the deviation of local hole density from the average $\overline{\rho}$ (which is set by gate voltage), and $C(\bm r)=e^2/(4\pi\epsilon|\bm r|)$ is the Coulomb potential with dielectric constant $\epsilon$, which controls the interaction strength. We approximate the Coulomb interaction $H_{C}$ by the mean-field Hatree potential $V_{H}$ self-consistently
\begin{equation}
V_{H}(\bm r)=V(\bm r)+\int C(\bm r-\bm r')\langle\delta\rho(\bm r')\rangle d^2\bm r',
\end{equation}
and $\langle\dots\rangle$ denotes the expectation value in mean-field ground state.
As we assume the Hartree potential preserves all symmetries, $V_{H}$ can be written as Fourier series similar to Eq. (\ref{eq_v}), and the Coulomb interaction only renormalizes the band structure \cite{guinea2018electrostatic,cea2019pinning}.

\begin{figure}[t]
\includegraphics[width=0.9\linewidth]{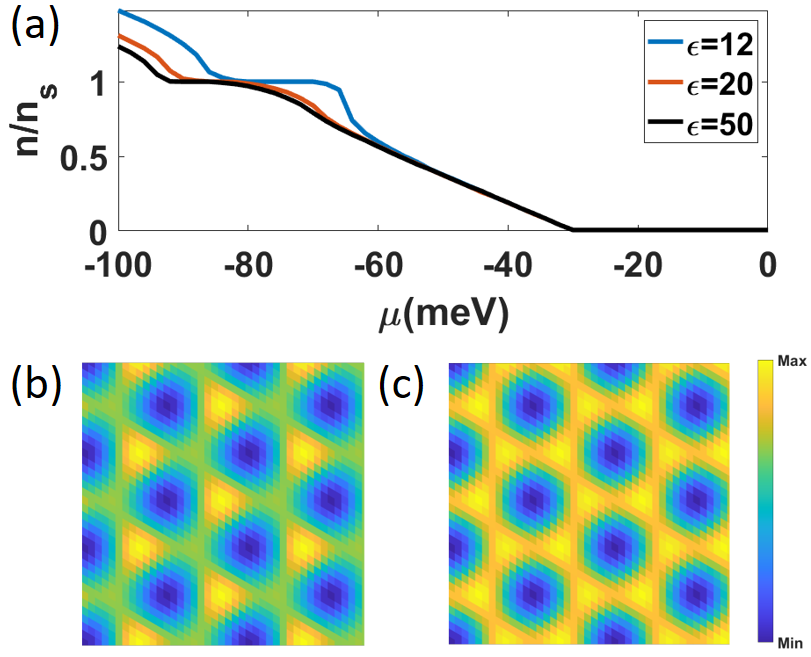}
\caption{(a) Filling factor $n/n_s$ as a function of chemical potential $\mu$ in WSe$_2$/WS$_2$ system at $\theta=3^\circ$ with $\epsilon=12, 20, 50$. (b) and (c) are charge distribution at half filling when $\epsilon=50$ and 12.}
\label{fig4}
\end{figure}

In Fig. \ref{fig4}a we plot the renormalized filling factor $n/n_s$ as a function of chemical potential $\mu$ in WSe$_2$/WS$_2$ heterobilayer at twist angle $\theta=3^\circ$ with different dielectric constants. At low fillings, the charge is always localized at MM spots. As we increase the filling, more holes will be accumulated and the repulsive interaction renormalizes the charge distribution to make it more homogeneous. Near half-filling $n=\frac{1}{2}n_s$, when the interaction is weak, the charge distribution remains at MM spots as shown in Fig. \ref{fig4}b.
When the interaction is strong, charge transfer from MM to MX spots occurs and the corresponding charge distribution is shown in Fig. \ref{fig4}c.

In real space, the interaction-induced, filling-dependent charge transfer leads to a significant change of charge distribution on the scale of 10 nm, which can be detected by scanning tunneling spectroscopy (STS). In the energy domain, charge transfer affects band structure on the scale of 10 meV, which may be detected in angle-resolved photoemission spectroscopy and optical measurement of the exciton spectrum.

To go beyond the mean-field approximation, in the following we provide a theoretical description of charge transfer physics in TMD heterobilayer with a sufficiently large $L_{\rm M}$, where the moir\'e bandwidth $W$ is small compared to the moir\'e potential $V_0$.

\emph{Tight-Binding Limit}---
We first introduce the moir\'e kinetic energy as $E_0\equiv (2mL_{\rm M}^{2})^{-1}$, which increases with the twist angle as $E_0\propto({\theta^2+\delta^2})$.
When moir\'e potential is weak compared with kinetic term $V_0\ll E_0$ (nearly free limit), the first and second bands have a negative indirect gap (e.g. Fig. \ref{fig2}b).
When moir\'e potential is much stronger than kinetic term $V_0/E_0\gg 1$ (tight-binding limit), moir\'e bands become flat compared with band gaps $W\ll\Delta$ (e.g. Fig. \ref{fig2}d).
All untwisted heterobilayers listed in Table I belong to the tight-binding limit as shown in Fig. \ref{fig3}.

In the tight-binding limit, each potential minimum traps a set of local Wannier orbitals.
The lowest-energy one is $s$-orbital, and the next is $(p_x, p_y)$ doublet. The first moir\'e band is predominantly formed by $s$ orbitals at MM spots, which are global potential minimum in the parameter range of interest $\phi\in(0,\frac{1}{3}\pi)$. The character of the second moir\'e band depends on $\phi$. For $\phi\in(0,\frac{1}{6}\pi)$, it comes from $p$ orbitals at MM spots. For $\phi\in(\frac{1}{6}\pi,\frac{1}{3}\pi)$ and in a wide range of $V_0$, it comes from $s$-orbitals at MX spots that are local potential minima (see Fig. \ref{fig2}). The energy difference between $s$ orbitals in MX and MM spots defines a charge transfer gap $\Delta_0 =\varepsilon_{MX}-\varepsilon_{MM}$.

By expanding $V(\bm r)$ around a potential minimum, we obtain the characteristic size of $s$-orbitals in MM and MX spots from the harmoinc approximation
\begin{eqnarray}
& & \psi(r, R) =\frac{1}{\sqrt{\pi} \xi} \exp \left(-\frac{|r-R|^{2}}{2 \xi^{2}}\right) \\
& & \xi_{MM}= (\cos\phi)^{-1/4} \xi_0 , \; \xi_{MX}=[\sin(\phi-\frac{1}{6}\pi)]^{-1/4} \xi_0 \nonumber \\
& & \xi_0 = ({4\pi^2 mV_0})^{-1/4} \sqrt{L_{\rm M}}
\end{eqnarray}
where $\xi_{MX}$ only applies to $\phi\in(\frac{1}{6}\pi,\frac{1}{3}\pi)$. It is important to note that for large $L_{\rm M}$, $\xi_i \propto \sqrt{L_{\rm M}}$ is parametrically smaller than the moir\'e period. Therefore, the local Coulomb repulsion is the largest interaction energy, given by
\begin{equation}
U_{i}=\frac{e^2}{4\sqrt{2\pi}\epsilon\xi_i}\propto L_{\rm M}^{-1/2}.
\end{equation}
with $i=MM, MX$. In contrast, the interaction between nearest neighbors $V'$ is proportional to $1/L_{\rm M}$ and hence parametrically smaller than $U$.

In Fig. S3 of the Supplementary Material we plot the bandwidth $W$ of the first moir\'e band and interaction energies $U,V'$ of WSe$_2$/WS$_2$ at different twist angles.
While $U, V'$ decrease with $L_{\rm M}$ in a power-law manner, $W$ is exponentially small in the tight-binding regime.
For untwisted WSe$_2$/WS$_2$, we find $L_{\rm M}=8.2$ nm, $\xi_{MM}=2.3$ nm, $\xi_{MX}=3.0$ nm and $U_{MM}=764/\epsilon$ meV, $U_{MX}=594/\epsilon$ meV, $W$=8 meV, $\Delta E_K=18 $ meV, $V'=302/\epsilon$ meV.

Depending on the relative strengths of interaction energy, bandwidth, and charge transfer gap, we find three phases at half-filling.

(I) Metal: $U\ll W$. The system is gapless. Under doping, additional charges are mainly localized at MM spots with $s$-orbital symmetry.

(II) Mott insulator: $\Delta>U\gg W$. The insulating ground state has one hole per MM spot, and the charge gap is $U$. When doped further, additional charges are mainly localized around MM spots. In this case, the triangular lattice Hubbard model is a good description \cite{wu2018hubbard}.

(III) Charge-transfer insulator: $U>\Delta\gg W$. The insulating ground state has one hole per MM spot, but the charge gap is $\Delta$. When further doped, additional charges are mainly localized at MX spots, thus resulting in charge transfer on moir\'e scale as the filling increases.

The insulating gap at half-filling inferred from thermal activation of resistivity is only around 10 meV \cite{regan2019optical,tang2019wse2}, which is significantly smaller than the estimated on-site repulsion $U \sim 128$ meV assuming $\epsilon = 6$. (Note the distance from the sample to metallic gates is 20 nm so that screening has little effect on local repulsion $U$). However, the measured gap is comparable to the charge transfer gap $\Delta E_K\sim 18$ meV. We thus conclude that the insulating phase at half-filling in untwisted WSe$_2$/WS$_2$ is likely a charge-transfer insulator, rather than a Mott-Hubbard insulator.

In order to capture the physics of charge transfer between MM and MX spots, we introduce an extended Hubbard model on the honeycomb lattice:
\begin{eqnarray}\label{eq_sh}
H=\frac{\Delta}{2}\sum_{i}(-)^i c^{\dagger}_{i}c_{i}-t\sum_{\langle ij\rangle}(c^{\dagger}_{i}c_{j}+h.c.)+\sum_{ij}V_{ij}n_in_j,
\end{eqnarray}
where $c_{i}=\{c_{i\uparrow},c_{i\downarrow}\}^{\rm T}$ denotes $s$-orbital holes, $(-)^i=\pm$ for $i=$MX (MM) spots and $t$ denotes hopping. $V_{ij}$ is the Coulomb repulsion between $s$ orbitals at site $i$ and $j$, which includes both on-site repulsion $U_{MM},U_{MX}$, nearest-neighbor repulsion $V'$ and etc. When there is strong screening from the metallic gates, interactions decay rapidly with the distance between sites.

At temperatures below the charge gap, double occupancy is strongly suppressed by the on-site repulsion $U$. For the triangular lattice Hubbard model, the low-energy physics is described by the $t$-$J$ model\cite{spalek2007tj,lee2006doping} with hopping $t$ and antiferromagnetic superexchange interaction $J=4t^2/U>0$ between nearest neighbors. Magnetic susceptibility as a function of doping at various temperatures is shown in Supplementary Material.
For charge-transfer insulators such as WSe$_2$/WS$_2$ described by the honeycomb lattice model (\ref{eq_sh}), their magnetic properties call for future study.

The extended Hubbard model in honeycomb lattice has also been realized with cold atom in optical lattice\cite{jaksch1998cold,hofstetter2002high,greiner2002quantum,gemelke2009situ,soltan2011multi}. However, in cold atom systems, the lowest accessible temperature at present is on the order of hopping $t$, which is much higher than the exchange interaction $J$\cite{mazurenko2017cold}. In the TMD heterobilayer WSe$_2$/WS$_2$, the exchange energy is around $J\sim 0.05$ meV and the corresponding temperature ($\sim 1K$) is readily accessible for experiments.

In conclusion, we present a theory that maps the long period moir\'e system onto an atomic crystal with cation and anion and studies the correlated insulating behavior.
We find that the interplay between moir\'e potential and interaction strength gives rise to charge transfer insulator in heterobilayer TMD, and opens the possibility of novel electronic states upon doping.

\section*{Acknowledgment}
We thank Kin Fai Mak and Jie Shan for sharing their experimental results with us prior to publication, and Hiroki Isobe, Zhen Bi, Taige Wang, and Feng Wang for helpful discussions. This work is supported by DOE Office of Basic Energy Sciences under Award DE-SC0018945, N.Y. was partly supported by the U.S. Department of Energy, Office of Science, Basic Energy Sciences, under Award Number DE-SC0020149.

\bibliography{ref}

\newpage

\appendix
\setcounter{figure}{0}
\renewcommand{\thefigure}{S\arabic{figure}}
\setcounter{equation}{0}
\renewcommand{\theequation}{S\arabic{equation}}
\setcounter{table}{0}
\renewcommand{\thetable}{S\arabic{table}}

\section{Commensurate structure}
The lattice constants for MoS$_2$, MoSe$_2$, WS$_2$ and WSe$_2$ are 3.18818, 3.31579, 3.18719 and 3.31698 Angstrom respectively\cite{mounet2018two}.
For bilayer with the same chalcogen atoms, the lattice mismatch is less than 0.1\%, While for bilayer with different chalcogen atoms, the lattice mismatch is around 4\%. To build a commensurate structure from two different monolayers with same type Bravais lattices, we consider a bilayer system, whose primitive vectors are denoted by $\{\bm a_1,\bm a_2\}$ and $\{\bm a'_1,\bm a'_2\}$ respectively. When the bilayer system is a commensurate superlattice with primitive vectors $\{\bm L_1,\bm L_2\}$, we have\cite{wu2018hubbard}
\begin{eqnarray}
&&
	\begin{pmatrix}
 \bm L_1\\
 \bm L_2
 \end{pmatrix}
 =
 M
 \begin{pmatrix}
 \bm a_1\\
 \bm a_2
 \end{pmatrix}
 =
 M'
 \begin{pmatrix}
 \bm a'_1\\
 \bm a'_2
 \end{pmatrix},\\
 &&M=
 \begin{pmatrix}
 m & n\\
 p & q
 \end{pmatrix},\quad
 M'=
 \begin{pmatrix}
 m' & n'\\
 p' & q'
 \end{pmatrix},
\end{eqnarray}
where $m,n,p,q,m',n',p',q'$ are eight integers. These integers are determined by lattice mismatch between two layers (i.e. twist angle and strain) and also information of each layer (such as the anisotropy ratio). The most general way to determine these integers out of lattice information is through enumeration. We search all eight integers in a given range and compute corresponding lattice information for every given set of eight integers. When the calculated lattice information matches with the given one (up to some given precision), we then find the solution. In the following, however, we will discuss two special classes of superlattices where these eight integers have analytical solutions.
When the two monolayers with threefold rotations, i.e. they are all triangular lattices, the integer matrices $M,M'$ have to be conformal (i.e. a scalar times a rotation). Without loss of generality, let us assume $\bm a_{1,2}=a(1/2,\pm\sqrt{3}/2)$ and $\bm a'_{1,2}=a'R(\theta)(1/2,\pm\sqrt{3}/2)$ with $a'\geqslant a$, namely the unprimed layer is not rotated while the primed layer is rotated by angle $\theta$ along out-of-plane direction. Then the conformal matrices $M,M'$ will have the following form
\begin{eqnarray}
 M=N
 \begin{pmatrix}
 \cos\phi -\frac{1}{\sqrt{3}}\sin\phi &  -\frac{2}{\sqrt{3}}\sin\phi\\
 \frac{2}{\sqrt{3}}\sin\phi & \cos\phi +\frac{1}{\sqrt{3}}\sin\phi
 \end{pmatrix},\\
 M'=N'
 \begin{pmatrix}
 \cos\phi' -\frac{1}{\sqrt{3}}\sin\phi' &  -\frac{2}{\sqrt{3}}\sin\phi'\\
 \frac{2}{\sqrt{3}}\sin\phi' & \cos\phi +\frac{1}{\sqrt{3}}\sin\phi'
 \end{pmatrix},
\end{eqnarray}
where $N,N'$ are positive integers and $\phi,\phi'$ are angles. Since each one of $M,M'$ is effectively described by two parameters, we can write
\begin{equation}
 M=
 \begin{pmatrix}
 m & n\\
 -n & m-n
 \end{pmatrix},\quad
 M'=
 \begin{pmatrix}
 m' & n'\\
 -n' & m'-n'
 \end{pmatrix}.
\end{equation}

Furthermore, we want to consider first-order moir\'e pattern for simplicity, which is defined in terms of reciprocal vectors. Assume the reciprocal lattice vectors of two layers and the superlattice are $\{\bm G_1,\bm G_2\}$, $\{\bm G'_1,\bm G'_2\}$ and $\{\bm g_1,\bm g_2\}$ respectively, then the first-order moir\'e pattern is defined by the conditions
\begin{equation}
 \bm g_{i}=\bm G_{i}-\bm G'_{i}.
\end{equation}
With this condition, it is found that $m-m'=n-n'=1$, and hence the twist angle and lattice constant ratio between two layers are
\begin{eqnarray}
 \theta =\arctan\left(\frac{\sqrt{3}(m-n)}{2(m^2+n^2-mn)-m-n}\right),
 \\
 r\equiv\frac{a'}{a}=\sqrt{\frac{m^2+n^2-mn}{m^2+n^2-mn-m-n+1}}
\end{eqnarray}
and the superlattice vectors are
\begin{eqnarray}
 \bm L_1=m\bm a_1+n\bm a_2,\quad
 \bm L_2=-n\bm a_1+(m-n)\bm a_2,\\
 L\equiv|\bm L_i|=\frac{a}{\sqrt{1-2r\cos\theta +r^2}}.
\end{eqnarray}

Here for TMDs with less than 0.1\% lattice mismatch, we construct the commensurate moir\'e superlattice via twist the top layer with angles
$\theta= 21.78^\circ, 13.1^\circ, 9.43^\circ, 6.59^\circ$ and $5.49^\circ$.
While for TMDs with lattice mismatch around 4\%, the calculated twist angels are $\theta= 16.31^\circ, 7.31^\circ, 6.39^\circ, 5.68^\circ$ and $4.72^\circ$.

\section{Details of the ab-initio calculation and parameter fitting}
For MoSe$_2$/WSe$_2$ and MoS$_2$/WS$_2$ with nearly identical lattice constant, we fit the continue model parameters from the energy shift of band maxima from the relative shift of monolayer unit cells\cite{wu2018hubbard}, as shown in Table. \ref{T}. We note that the fitting from large-scale DFT as we have done in the main text would give different parameters, and leave the more realistic treatment such as lattice relaxation at the long moir\'e wavelength limit to future study.

\begin{figure}[t]
\includegraphics[width=1.0\linewidth]{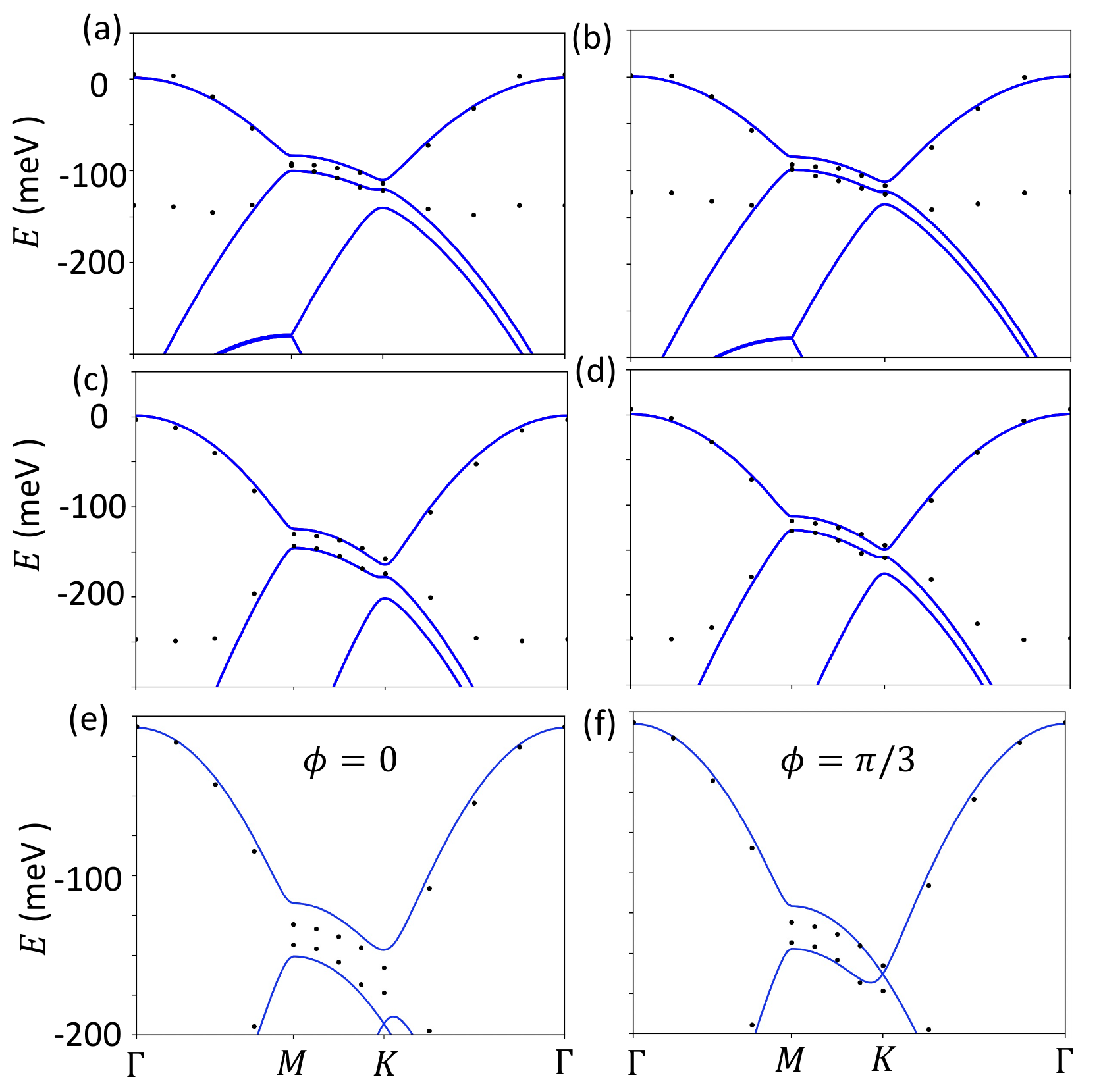}
\caption{Continuum model fitting of DFT band structure at twist angle $\theta=5.68^\circ$ of (a)MoSe$_2$/MoS$_2$, (b)MoSe$_2$/WS$_2$, (c)WSe$_2$/MoS$_2$
and (d)WSe$_2$/WS$_2$. 
DFT band structure of WSe$_2$/WS$_2$ at twist angle $\theta=5.68^\circ$ and continuum model fitting with $V=15$ meV and (e) $\phi=0$, (f) $\phi=\pi/3$.}
\label{sup1}
\end{figure}

For another four systems with lattice mismatch 4\%, we fit the parameters of periodic potential and its phase factor directly from DFT band structures at various commensurate structures with different twist angles, a further calculation of DFT charge density distribution is performed to fix sign of the phase factor.
We note that bilayer structures with twist angle $\theta=5.68^\circ$ fall into the gauge that $K$ pockets of monolayer unit cell fold to $\Gamma$ point of moir\'e BZ. The band structures of various TMD heterobilayers with unequal monolayer lattice constants are summarized in Fig. \ref{sup1}. Note the energy of $\Gamma$ pockets in MoSe$_2$ is only lower by 120 meV compared with K pockets in MoSe$_2$, which gives rise to the relatively flat second Moir\'e  band around $\Gamma$  point in moir\'e BZ in Fig. \ref{sup1} (a, b).

Since the charge transfer physics is dependent on the bandwidth $W$ and charge transfer gap $\Delta$, we choose the bandwidth of the first moir\'e band and the bandgap between first and second moir\'e band as the essential criteria for the fitting of continuum model parameters. As expected, the potential strength $V_0$ determines the moir\'e bandwidth, while the potential phase factor $\phi$ determines the bandgap at $K$ point in the moir\'e BZ. As shown in Fig. \ref{sup1}, we also list the band structure fitting with different phase factor $\phi$ for WSe$_2$/WS$_2$ heterobilayer, which shows a large bandgap in $K$ at $\phi=0$ and a degenerate $K$ point at $\phi=\pi/3$ for the first and second moir\'e band.

\begin{table}\setlength{\tabcolsep}{7pt}
\centering
\begin{tabular}{c c c c c c}
\hline\hline
System & $\delta$ & $\displaystyle \Delta E_g$ & $\displaystyle V_0$ & $\displaystyle \phi$ & $\displaystyle E_{0}^{\rm min}$ \\[0.3cm]
\hline
WSe$_2$/MoSe$_2$ & $<0.1\%$ & 370 & 8.14 & -4$^\circ$ & $<10^{-3}$\\[0.2cm]
WS$_2$/MoS$_2$ & $<0.1\%$ & 360 & 6.52 & -13$^\circ$ & $<10^{-3}$\\[0.2cm]
\hline\hline
\end{tabular}
\caption{Summary of heterobilayer TMD. Here $\delta$ is the lattice constant mismatch with respect to the bottom layer, $\Delta E_g$ is the band offset, $V_0$ and $\phi$ are parameters of moir\'e potential and $E_{0}^{\rm min}=\delta^2/(2ma^2)$ is the moir\'e kinetic energy at zero twist. All energies are in unit of meV.}
\label{T}
\end{table}

\section{Fock term and Details of charge transfer}
The full mean-field treatment of the Coulomb interaction also includes Fock decomposition $\delta\rho(\bm r)\delta\rho(\bm r')\to \psi^{\dagger}(\bm r)\psi(\bm r')\langle\psi(\bm r')\psi^{\dagger}(\bm r)\rangle$, which results in the mean-field Hamiltonian $H_{\rm MF}=\int\psi^{\dagger}(\bm r)g(\bm r-\bm r')\psi(\bm r')d^2\bm rd^2\bm r'$ with renormalized propagator
$g(\bm r)=g_{0}(\bm r)+C(\bm r)\langle\psi(\bm 0)\psi^{\dagger}(\bm r)\rangle$,
where $g_0(\bm r)$ is the bare propagator.
We expect that the moir\'e potential and hence charge transfer physics discussed in this work would not be affected by Fock term too much.

\begin{figure}[ht]
\includegraphics[width=1\linewidth]{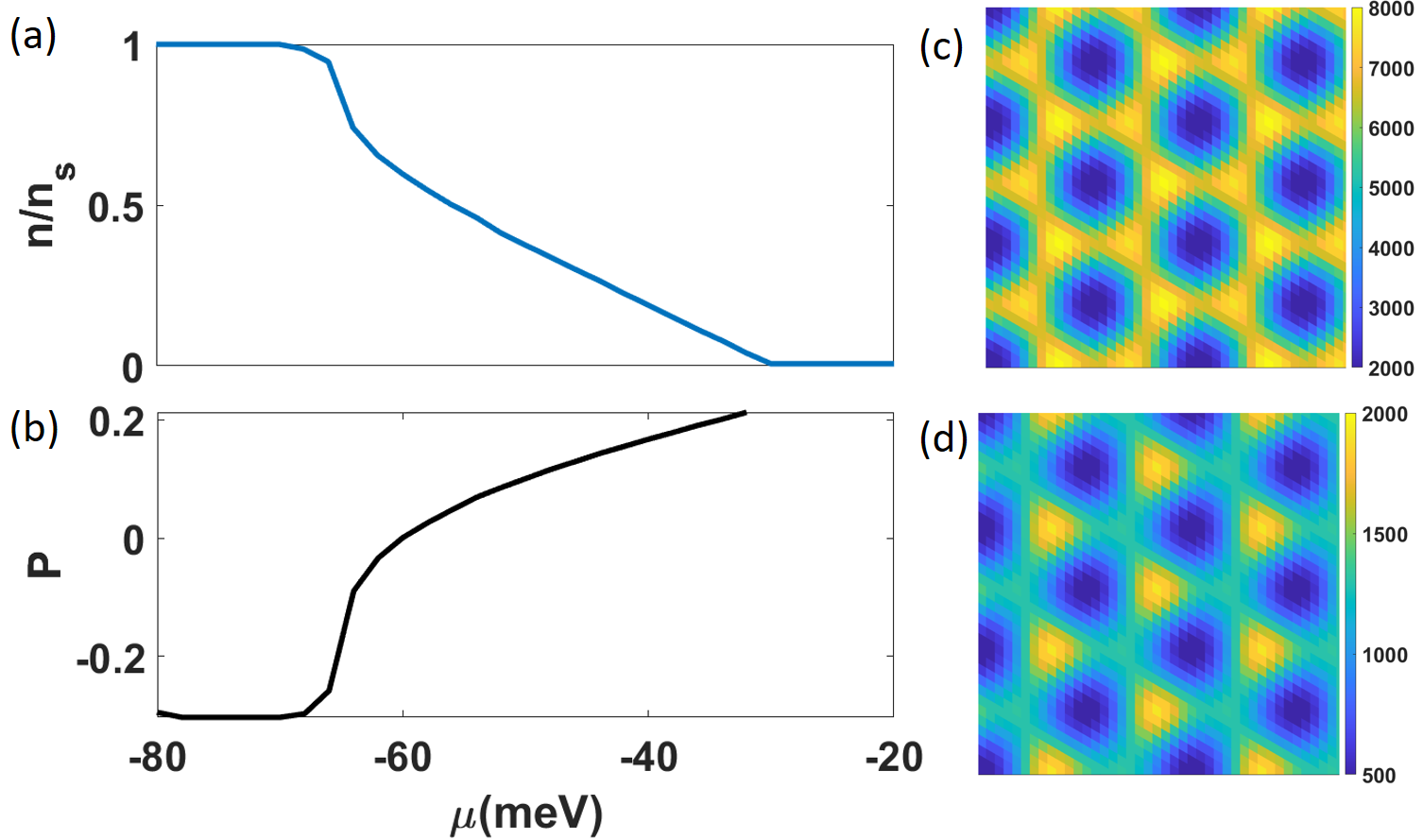}
\caption{Charge transfer in WSe$_2$/WS$_2$ system at twist angle $\theta=3^\circ$ with dielectric constants $\epsilon=12$. (a) and (b) are filling factor $n/n_s$ and charge imbalance parameter $P$ as functions of chemical potential $\mu$, (c) and (d) are charge distribution at different fillings $n=0.5n_s$ and $n=0.1n_s$ respectively. Units of colorbars are arbitrary.}
\label{sup2}
\end{figure}

We can introduce a dimensionless quantity to describe the charge imbalance between MM and MX spots
\begin{equation}
 P=\frac{n_{\rm MM}-n_{\rm MX}}{n_{\rm MM}+n_{\rm MX}},
\end{equation}

and plot it as a function of chemical potential $\mu$ together with filling factor, as shown in Fig. \ref{sup2}a and b. As we can see, at low filling, the charge is mainly concentrated at MM spots $P>0$, when doping increases, MM and MX spots first become balanced in terms of charge distribution $P\to 0$, and then MX spots can have more charges than MM spots $P<0$ since repulsion at MX spots is weaker than MM.
As explicit examples, we also plot charge distribution at different fillings $n=0.5n_s$ and $n=0.1n_s$ in Fig. \ref{sup2} c and d respectively.

\section{Band structures, Coulomb interactions, and tight-binding models}

\begin{figure}[!htb]
\includegraphics[width=1\linewidth]{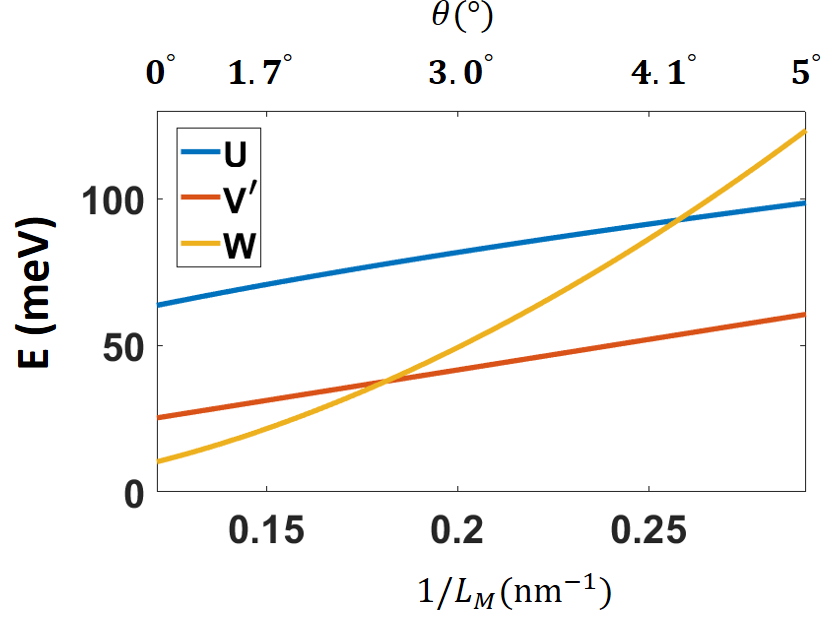}
\caption{Onsite Coulomb repulsion $U$, nearest-neighbor repulsion $V'$ and bandwidth $W$ of WSe$_2$/WS$_2$ as functions of moir\'e wavelength $L_{\rm M}$ (bottom axis) and twist angle $\theta$ (top axis) when dielectric constant is chosen as $\epsilon=12$.
}
\label{sup3}
\end{figure}

In Fig. \ref{sup3}, we plot onsite repulsion $U$, nearest-neighbor repulsion $V'$ and bandwidth $W$ of WSe$_2$/WS$_2$ at different twist angles. All three energy scales decrease with decreasing angle $\theta$ (or equivalently increasing moir\'e wavelength $L_{\rm M}$), but $W$ decreases faster than interactions.

As shown in Fig. 2, there are two regimes in the phase space spanned by moir\'e potential strength $V_0/E_0$ and moir\'e phase $\phi$, where the second moir\'e bands are qualitatively different.

Triangular regime where the second and third moir\'e bands are from $(p_x,p_y)$ orbitals at MM spots and the corresponding tight-binding model is
\begin{eqnarray}
 H_{\triangleright}=\sum_{i}\bm c^{\dagger}_{i}\hat{\Delta}\bm c_{i}-\sum_{\langle ij\rangle}(\bm c^{\dagger}_{i}\hat{T}\bm c_{j}+h.c.),
\end{eqnarray}
where $\bm c=\{s,p_{x},p_{y}\}^{\rm T}$ denotes $s$- and $p_{x,y}$-orbital holes, and $i$ is MM spot. $\hat{\Delta}=$diag$(\varepsilon_{s},\varepsilon_{p},\varepsilon_{p})$ denotes onsite energy matrix, where $\varepsilon_{s},\varepsilon_{p}$ denote the energy of $s$ and $(p_x,p_y)$ orbitals at MM spot respectively.

Honeycomb regime where the second moir\'e band is from $s$ orbitals at MX spots and the corresponding tight-binding model is
\begin{eqnarray}
 H_{\hexagon}=\frac{\Delta}{2}\sum_{i}(-)^i c^{\dagger}_{i}c_{i}-t\sum_{\langle ij\rangle}(c^{\dagger}_{i}c_{j}+h.c.),
\end{eqnarray}
where $c_{i}=\{c_{i\uparrow},c_{i\downarrow}\}^{\rm T}$ denotes $s$-orbital holes, $(-)^i=\pm$ for $i=$MX (MM) spots and $t$ denotes hopping.
And $\Delta =\varepsilon_{MX}-\varepsilon_{MM}$ denotes the charge-transfer gap from MM to MX spots.

\section{Magnetic Properties}
We also study the magnetic properties of $t$-$J$ model in the triangular lattice. We performed ED (exact diagonalization) \cite{bauer2011alps} on the triangular lattice $t-J$ model, and calculate temperature and filling dependent spin susceptibility.  

\begin{figure}[ht]
\includegraphics[width=1.0\linewidth]{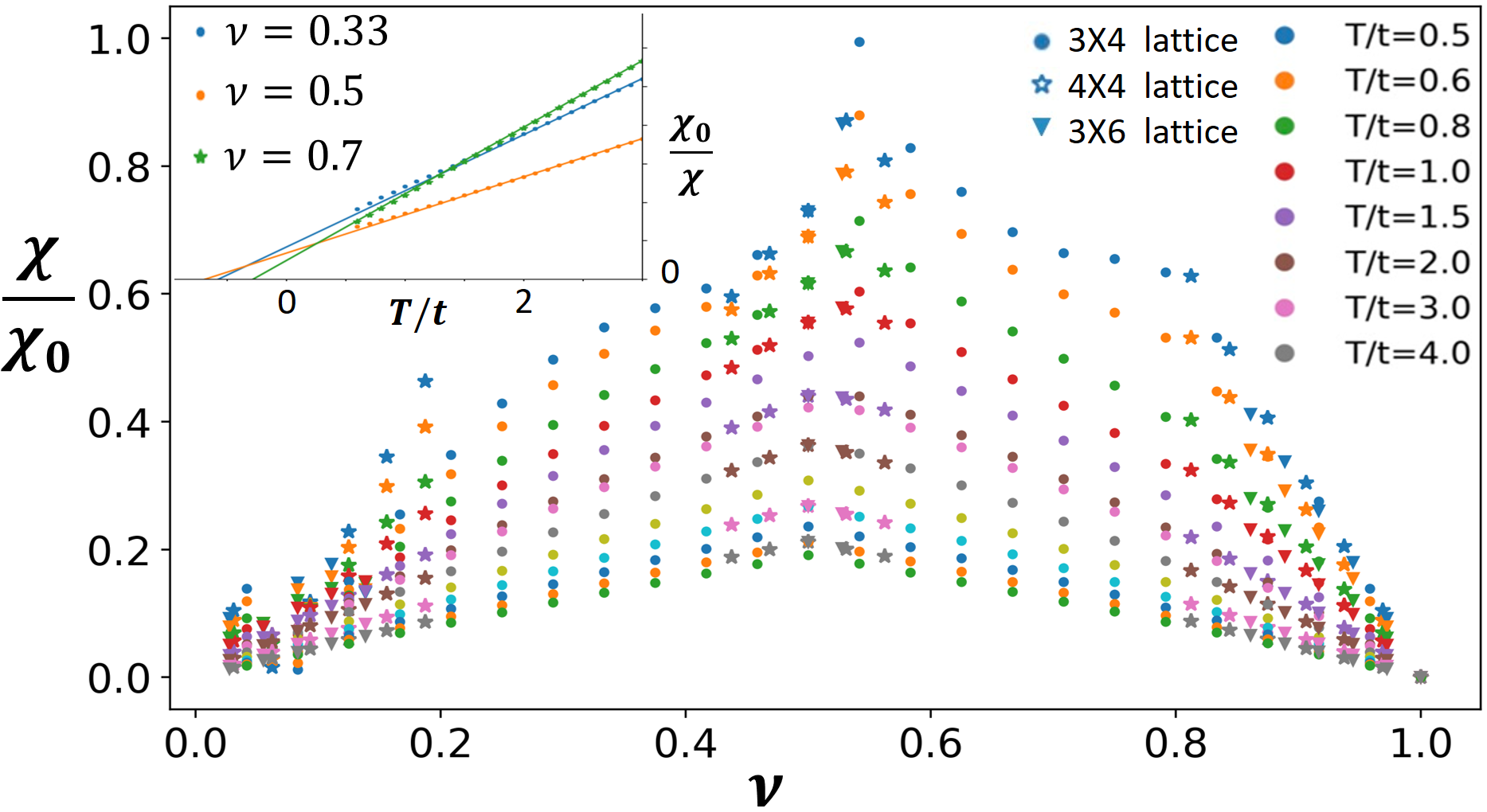}
\caption{
Uniform spin susceptibility of $t$-$J$ model at $J/t$=0.4 $ ({U}/{t}=10$). The peak is around $\nu=n/n_s=\frac{13}{24}$ for $0.5t<T<1.5t$, but at $\nu=\frac{1}{2}$ for $T>1.5t$. Here $\chi_0=t^{-1}$, and $t$ is around 1 meV.
Inset: Inverse spin susceptibility as a function of temperature, where dots are numerical results and solid lines are Curie-Weiss fit.
The Curie temperature $T_0$ is largest in magnitude at half-filling and gradually moves towards zero as the doping deviates from half-filling, indicating the transition from AFM to PM.
}
\label{sup4}
\end{figure}

From the Curie-Weiss law plot of susceptibilities in Fig. \ref{sup4} inset, we found that Curie temperature $T_0$ moved towards zero when we increased the filling factor higher than half, indicating that antiferromagnetic correlation is reduced by doping.
As shown in Fig. \ref{sup4}, in a wide temperature range, spin susceptibility is a non-monotonic function of filling factor $\nu=n/n_s$ with a maximal peak at optimal filling $\nu(T)$. Among them, at higher temperature $T\gg t$ the peak locates exactly at the half-filling $\nu=\frac{1}{2}$, while at lower temperature $T\lesssim t$, the susceptibility peak is shifted to above half-filling $\nu>\frac{1}{2}$, in agreement with the result from high-temperature expansion study\cite{koretsune2002resonating}.
These findings are consistent with the spin susceptibility of WSe$_2$/WS$_2$ heterobilayers inferred from optical spectroscopy under the magnetic field
 \cite{tang2019wse2}. 

\end{document}